\documentstyle[epsbox]{article}

\date{}
\input epsf
\title{\bf Distribution of Instanton and Monopole Clustering}
\author{\underline{M. Fukushima},
A. Tanaka, S. Sasaki, H. Suganuma, H. Toki and  D. Diakonov  $^{\rm a}$,\\ 
\\
\small \it Research Center for Nuclear Physics (RCNP), Osaka University,\\
\small \it 10-1 Mihogaoka, Ibaraki, Osaka 567, Japan\\
\\
\small $^{\rm a}$\it Petersburg Nuclear Physics Institute (PNPI),\\
\small \it Gatchina, St.Petersburg 188350, Russia}
\begin{document}

\maketitle

\begin{abstract}
We study the relation between the instanton distribution and the monopole 
loop length in the SU(2) gauge theory with the abelian gauge fixing. We 
measure the monopole current from the multi-instanton ensemble on the 
$16^4$ lattice using the maximally abelian gauge. When the instanton 
density is dilute, there appear only small monopole loops. On the other 
hand, in the dense case, there appears one very long monopole loop, which 
is responsible for the confinement property, in each gauge configuration. 
We find a clear monopole clustering in the histogram of the monopole loop 
length from 240 gauge configurations. \\
\end{abstract}

\section{Topological Objects in QCD}
In QCD, there are two non-trivial topological objects, which are important 
for understanding non-perturbative properties. QCD is reduced to an abelian 
gauge theory with QCD-monopole after performing the abelian gauge fixing [1]. 
The monopole appears corresponding to the non-trivial homotopy group, 
$\pi_{2}(SU(N_{c})/U(1)^{N_{c}-1})=Z_{\infty}^{N_{c}-1}$, and monopole 
condensation plays an essential role on color confinement and chiral symmetry 
breaking [2,3]. On the other hand, an instanton appears as a classical 
non-trivial solution in Euclidean 4-space corresponding to the homotopy group, 
$\pi_{3}(SU(N_{c}))=Z_{\infty}$ [4]. Also, instantons are very important for 
the phenomena related to the $U_{A}(1)$ anomaly and chiral symmetry breaking. 
Until now, however, there has been no evidence the instanton has anything to 
do with color confinement. \\

Thus, it seems that instantons and monopoles are belonging completely 
different sectors of physics, and are hardly related to one another. However, 
both of them should be essential key for non-perturbative QCD. Therefore, we 
investigate the relation between instantons and monopoles in terms of the 
color confinement mechanism.\\

\section{Strong Correlation between Instanton and Monopole}
Recently, the strong correlation between instantons and monopoles has been 
found both in the analytical framework and the lattice QCD [5-11]. There are 
two remarkable points. First, from the analytical studies [6,8-11], each 
instanton seems to accompany a monopole trajectory near its center. Such a 
correlation may lead to the linearity on the relation between the instanton 
number and the total monopole loop length, which is clearly observed in the 
lattice QCD [11]. Second, apart from the centers of instantons, monopole 
trajectories are very unstable against small fluctuation on the instanton's 
size and location [8,11]. Therefore, these fluctuations tend to combine 
isolated monopole trajectories into one longer trajectory. We then expect 
that monopole trajectories would combine neighboring instantons as the 
instanton density increases. Due to the above two facts, there would appear 
very long and complicated monopole trajectories in the non-perturbative QCD 
vacuum characterized by the dense topological pseudoparticles (instantons 
and anti-instantons). \\

The monopole clustering is observed in the lattice QCD simulation [11]. In 
the confinement phase, which includes many instantons, there appears one 
very long monopole loop which covers the entire lattice space in each gauge 
configuration [12]. This long and complicated monopole loop is a signal of  
monopole condensation, which is responsible for color confinement [13]. On 
the other hand, in the deconfinement phase, there appear only small monopole 
loops, which would not contribute to the confinement force. \\

We study the multi-instanton system in terms of the monopole clustering in 
order to understand the essence of the non-perturbative QCD vacuum.\\

\section{Multi-Instanton Configuration}
The multi-instanton ensemble is characterized by the density and the size 
distribution of instantons [14]. It has been analytically shown that the 
linear confinement potential can be obtained if the instanton size 
distribution falls off as $1/\rho^{3}$ at large $\rho$. While, the ordinary 
instanton liquid models suggest that the distribution behaves as $1/\rho^{5}$ 
at the large size. As for the small size, the distribution has to follow the 
one loop result, $f(\rho )\sim \rho^{b-5}$ with $b=11N_{c}/3$. Therefore, we 
adopt the size distribution as\\

\begin{equation}
f(\rho )=\frac{1}{(\frac{\rho}{\rho_{1}})^{\nu }+(\frac{\rho_{2}}{\rho})^{b-5}}
\end{equation}
with size parameters $\rho_{1}$ and $\rho_{2}$, which should satisfy the 
normalization condition, $\int^{\infty }_{0} d\rho f(\rho )=1$. The maximum 
of the distribution is fixed to the standard probable size $\rho_{0}$. We 
calculate for the two cases with $\nu =3,5$ for the large size distribution.\\

The gauge configuration of an instanton with the size $\rho$ and the center 
$z$ in the singular gauge is expressed as \\

\begin{equation}
A_{\mu }^{I}(x;z,\rho,O)=
\frac {i\tau^{a} \rho^{2}O^{ab} \bar{\eta }^{b}_{\mu \nu}(x-z)_{\nu}}
{(x-z)^{2}[(x-z)^{2}+\rho^{2}]},
\end{equation}
where $O$ denotes the SU(2) color orientation matrix and 
$\bar{\eta }^{a}_{\mu \nu } $ is the 't Hooft symbol. For an anti-instanton 
$A_{\mu }^{\bar{I}}$, one has to replace the $\bar{\eta }$ symbol with 
$\eta ^{a}_{\mu \nu } $. \\

The multi-instanton configurations are approximated as the sum of instanton 
and anti-instanton solutions, \\

\begin{eqnarray}
A_{\mu }(x)=\sum_{k} A_{\mu }^{I}(x;z_{k},\rho_{k},O_{k}) 
+\sum_{k} A_{\mu }^{\bar{I}}(x;z_{k},\rho_{k},O_{k}).
\end{eqnarray}   

We generate ensemble of pseudoparticles with random color orientations $O_{k}$ 
and centers $z_{k}$. The instanton sizes $\rho_{k}$ are randomly taken 
according to Eq.(1). These procedures are performed in the continuum theory. 
We then introduce a lattice on this gauge configuration and define the link 
variables, $U_{\mu }={\rm exp}(iaA_{\mu })$. We apply the maximally abelian 
gauge fixing [8,10-12], which maximizes 
$R=\sum_{\mu,s}{\rm Tr}[U_{\mu}(s) \tau^{3} U^{\dagger}_{\mu}(s) \tau^{3}]$, 
in order to extract monopole trajectories in the multi-instanton ensemble. 
We measure the monopole loop lengths and make the histograms of the monopole 
loop length.  \\

\section{Numerical Results}
We take a $16^{4}$ lattice with the lattice spacing $a=0.15{\rm fm}$, which 
means that the total volume is equal to $V=(2.4{\rm fm})^{4}$. The probable 
instanton size is fixed as $\rho_{0}=0.4{\rm fm}$ in our calculation. We show 
two typical cases with the total pseudoparticle number $N=20$ and $60$, which 
correspond to the density $(N/V)^{\frac{1}{4}}=174$ and $228$ ${\rm MeV}$, 
respectively. Fig.1 shows the histograms of monopole loop lengths for two 
density cases with $\nu =3$. At low instanton density (Fig.1(a)), only 
relatively short monopole loops remain. At high density (Fig.1(b)), there 
appears one very long monopole loop in each gauge configuration. As the 
density increases, a monopole trajectory tends to combine the overlapping 
pseudoparticles, and this mechanism generates one very long and highly 
complicated monopole loop. The appearance of the monopole clustering can 
be interpreted as a Kosterlitz-Thouless-type phase transition [13]. The 
critical density is found to be about $200 {\rm MeV}$ in the $\nu =3$ case. 
For the $\nu =5$ case, the similar tendency is obtained qualitatively, 
although the critical density is a little higher.\\

\section{Discussion and Concluding Remarks}
Finally, we compare our results with those of the true SU(2) lattice QCD 
with $16^{3}\times 4$ at different temperatures $(\beta =2.2$ and $2.35)$ 
[11]. The monopole clustering at the high density case (Fig.1(b)) resembles 
that in the confinement phase $(\beta =2.2)$, where the instanton density 
is relatively high as suggested in lattice QCD. On the other hand, the 
disappearance of the long monopole loop at the dilute case (Fig.1(a)) 
is similar to the result obtained in the deconfinement phase $(\beta =2.35)$. 
As the temperature increases, the instanton density is largely reduced [11] 
due to the instanton and anti-instanton pair annihilation. Therefore, each 
instanton becomes isolated, and the monopole loop originating from the 
instanton tends to be localized around each instanton. As a result, there 
remain only small monopole loops at high temperature. \\

Thus, the instanton density would play an essential role on the deconfinement 
phase transition through the monopole clustering, which is a signal of 
monopole condensation [13] \\

As interesting related subjects, we are now studying the local correlation 
between instantons and monopoles, and are calculating the Wilson loop in 
order to clarify the role of these topological objects on the color 
confinement mechanism directly. \\

\begin{figure}[htb]
\begin{center}
\epsfile{file=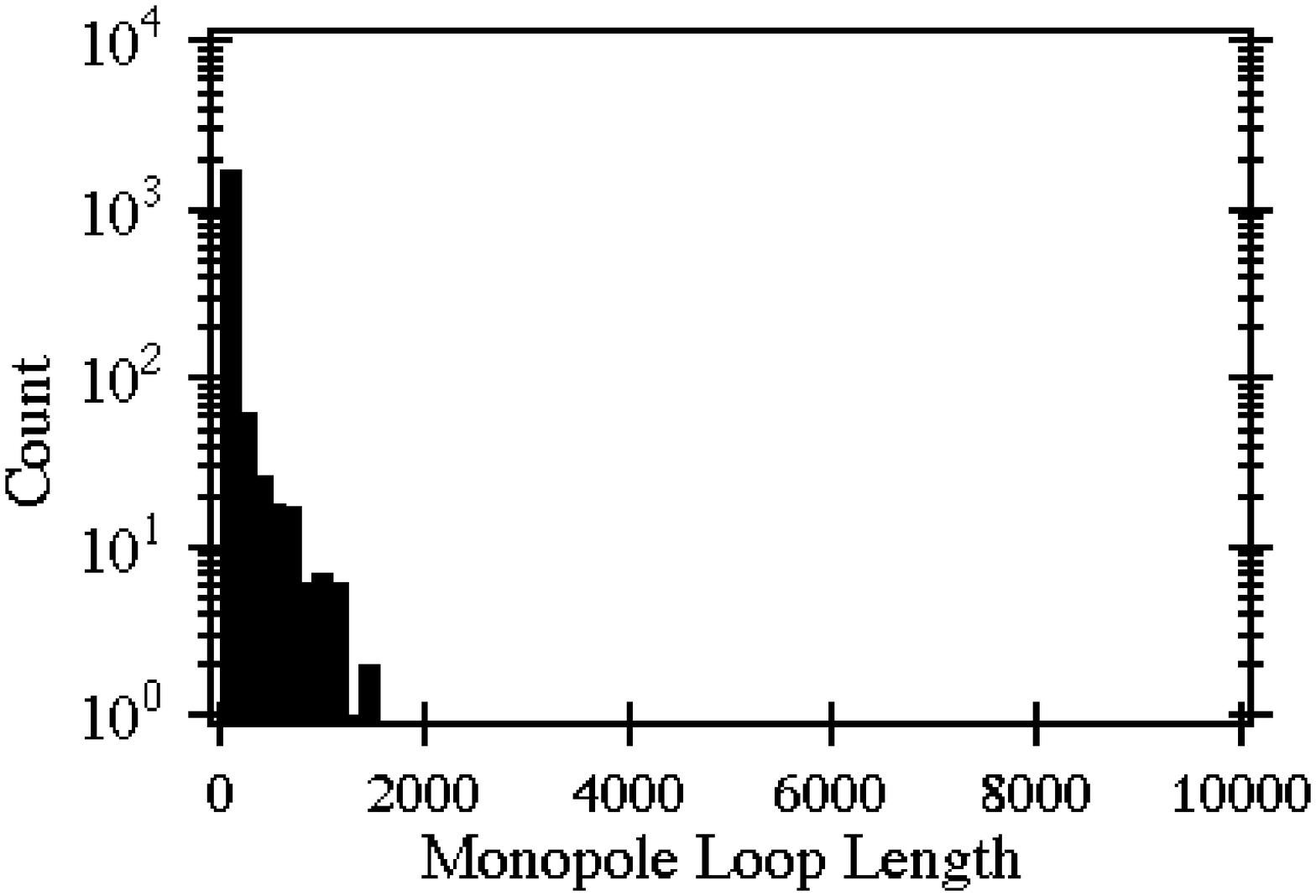,height=5.8cm}
\end{center}

{
\small
\noindent
\hspace{5.0cm}Fig.1(a):  dilute case  $[ \hspace{0.1cm}(N/V)^{1/4}=174 
{\rm MeV}\hspace{0.1cm}]$ \\
}
\end{figure}

\vspace{-0.5cm}

\begin{figure}[htb]
\begin{center}
\epsfile{file=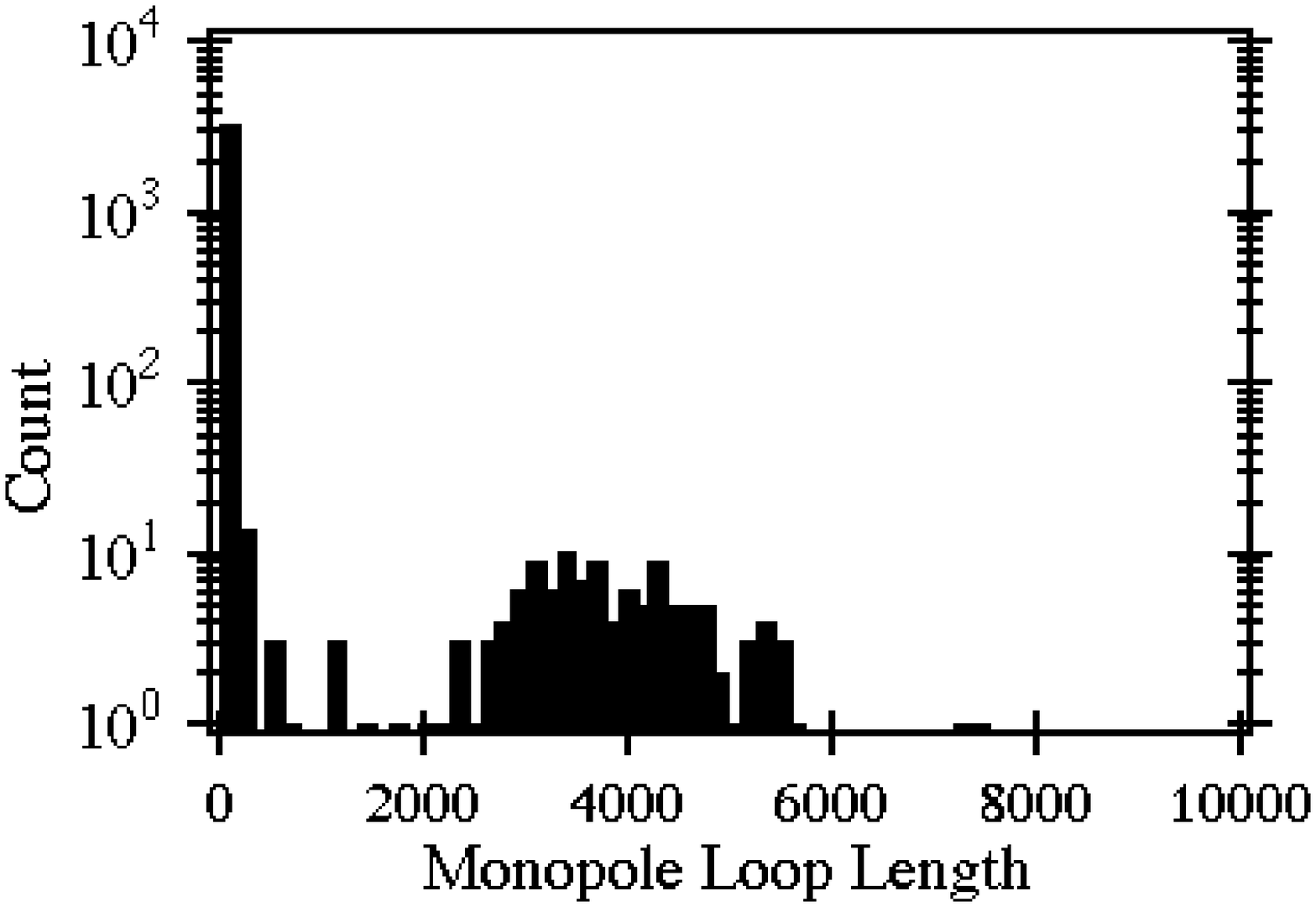,height=5.8cm}
\end{center}

{
\small
\noindent
\hspace{5.0cm} Fig.1(b):  dense case $[ \hspace{0.1cm}(N/V)^{1/4}=228 
{\rm MeV}\hspace{0.1cm}]$ \\
}
\end{figure}

\vspace{-0.2cm}

\hspace{1.0cm}Fig.1\hspace{0.4cm}Histograms of monopole loop lengths 
in the multi-instanton system; \\
\vspace{-0.4cm}
\noindent

\hspace{2.0cm} (a) dilute case and (b) dense case. \\


\begin{thebibliography}{9}
\bibitem{Hooft} G.~'t Hooft, Nucl.~Phys.~{\bf B190}~(1981)~455.
\bibitem{suga_4} H.~Suganuma, S.~Sasaki and H.~Toki, 
Nucl.~Phys.~{\bf B435}~(1995)~207.
\bibitem{miya_1} O.~Miyamura, Nucl.~Phys.~{\bf B}~(Proc.Suppl.)
{\bf 42}~(1995)~538.
\bibitem{Polya} A.~Belavin, A.~Polyakov, A.~Shvarts and Yu.~Tyupkin, 
Phys.~Lett.~{\bf 59B}~(1975)~85.
\bibitem{miya_2} O.~Miyamura and S.~Origuchi
in {\it Color Confinement and Hadrons}~(World Scientific,1995)~235.
\bibitem{suga_1} H.~Suganuma, H.~Ichie, S.~Sasaki and H.~Toki, 
in {\it Color Confinement and Hadrons}~(World Scientific,1995)~65.
\bibitem{markum} S.~Thurner, H.~Markum and W.~Sakuler, 
in {\it Color Confinement and Hadrons}~(World Scientific,1995)~77.
\bibitem{suga_2} H.~Suganuma, A.~Tanaka, S.~Sasaki and O.~Miyamura,
Nucl.~Phys.~{\bf B}~(Proc.Suppl.) {\bf 47}~(1996)~302.
\bibitem{cherno} M.~Chernodub and F.~Gubarev,~JETP~Lett.{\bf 62}~(1995)~100.
\bibitem{hart} A.~Hart and M.~Teper, Phys.~Lett.~{\bf B371}(1996)~261.
\bibitem{suga_4} H.~Suganuma, S.~Sasaki, H.~Ichie, H.~Toki and F.~Araki,
in {\it Frontier '96}~(World Scientific).
\bibitem{schier} F.~Brandstater, U.-J.~Wiese and G.~Schierholz, 
Phys.~Lett~{\bf B272}~(1991)~319.
\bibitem{ezawa} Z.~F.~Ezawa and A.~Iwazaki, Phys.~Rev.~{\bf D25}
(1982)~2681;~{\bf D26}~(1982)~631.
\bibitem{diakon1} D.~Diakonov and V.~Petrov, 
in {\it Nonperturbative Approaches to QCD}~(PNPI,1995)~239.

\end{thebibliography}
\end{document}